\documentclass[manuscript,screen]{acmart}

\usepackage{microtype}
\usepackage{booktabs}
\usepackage{amsmath}
\usepackage{graphicx}

\usepackage[capitalize,noabbrev]{cleveref}

\usepackage{siunitx}
\title[Pseudo-Label NCF for Sparse OHC Recommendation]{Pseudo-Label NCF for Sparse OHC Recommendation: Dual-Representation Learning and the Separability--Accuracy Trade-off}

\acmConference[RecSys 2026]{ACM Conference on Recommender Systems}{2026}{USA}
\acmYear{2026}

\author{Pronob Kumar Barman}
\affiliation{%
  \department{Department of Information Systems}
  \institution{University of Maryland, Baltimore County}
  \city{Baltimore}
  \state{Maryland}
  \country{USA}
}
\email{pbarman1@umbc.edu}

\author{Tera L.\ Reynolds}
\affiliation{%
  \department{Department of Information Systems}
  \institution{University of Maryland, Baltimore County}
  \city{Baltimore}
  \state{Maryland}
  \country{USA}
}
\email{reynoter@umbc.edu}

\author{James Foulds}
\affiliation{%
  \department{Department of Information Systems}
  \institution{University of Maryland, Baltimore County}
  \city{Baltimore}
  \state{Maryland}
  \country{USA}
}
\email{jfoulds@umbc.edu}

\renewcommand{\shortauthors}{Barman et al.}

\begin{document}

\begin{abstract}
Online Health Communities (OHCs) connect patients for peer support, but users face a discovery challenge when they have minimal prior interactions to guide personalization. We address recommendation under extreme interaction sparsity in a survey-driven setting where each user provides a 16-dimensional intake vector at registration and each support group has a structured feature profile. We augment three Neural Collaborative Filtering (NCF) architectures---Matrix Factorization (MF), Multi-Layer Perceptron (MLP), and NeuMF---with an auxiliary pseudo-label objective derived from survey--group feature alignment using cosine similarity mapped to $[0,1]$. This yields Pseudo-Label NCF (PL-NCF), which learns dual embedding spaces: main embeddings optimized for ranking, and PL-specific embeddings intended to capture semantic alignment between users and support groups.

On a dataset of 165 users, 498 support groups, and 498 observed memberships (three per user), we evaluate ranking quality (HR@5, NDCG@5) primarily under a leave-one-out protocol---one validation and one test interaction per user, yielding approximately one training positive per user---which best reflects our target cold-start regime, with a train-validation-test (70/15/15) split for supplementary analysis. We analyze embedding structure in the leave-one-out setting using spherical $k$-means silhouette scores computed in the original high-dimensional embedding space, alongside 2D t-SNE visualizations for qualitative inspection. Under leave-one-out evaluation, all three PL variants improve ranking: MLP-PL improves HR@5 from \qty{2.65}{\percent} to \qty{5.30}{\percent}, NeuMF-PL from \qty{4.46}{\percent} to \qty{5.18}{\percent}, and MF-PL from \qty{4.58}{\percent} to \qty{5.42}{\percent}. Critically, PL-specific embedding spaces exhibit higher cosine silhouette scores than baseline main embeddings under per-seed optimal clustering with $k \in \{3, 4, 5, 6, 7, 8, 10\}$: MF-PL improves from \num{0.0394} to \num{0.0684}; NeuMF-PL improves from \num{0.0263} to \num{0.0653}. We also observe a separability--accuracy trade-off: main-embedding clusterability is negatively correlated with ranking accuracy (Spearman $\rho \approx -0.38$ on leave-one-out; $\rho \approx -0.59$ on the train-validation-test split), suggesting that embeddings optimized for ranking may sacrifice interpretability. These findings demonstrate that survey-derived pseudo-labels can regularize sparse-data training for improved ranking, while producing interpretable task-specialized embedding spaces via dedicated PL-specific representations.
\end{abstract}



\maketitle

\section{Introduction}
Online Health Communities (OHCs) support patients managing chronic conditions by enabling peer connection, experiential knowledge exchange, and coping strategies \cite{barman2026understanding, ref1,ref2,ref3}. Yet the moment when support can matter most is often the moment when recommendation is technically hardest: at onboarding, users need timely discovery of relevant support groups, but platforms have little to no behavioral history to personalize recommendations. As OHC catalogs expand, this cold-start regime creates a discovery bottleneck in which users must choose among dozens or hundreds of groups with only minimal interaction signals, increasing the risk of poor initial matches and early disengagement.

We study support-group recommendation when the primary available input is a structured intake survey. Our setting differs from (i) group recommender systems that aggregate preferences for established groups \cite{ref4,ref5,ref6}, which assume existing member preferences, and (ii) content-based methods that mine forum text \cite{ref27}, which require post-hoc community analysis. Instead, we adopt a feature-driven approach where survey responses yield a compact 16-dimensional representation of user needs, and each group has a corresponding aggregated feature profile. This structure also provides an interpretable alignment signal: the cosine similarity between user and group features, which we term \emph{AlignFeatures}.

Neural Collaborative Filtering (NCF) models \cite{ref11} capture non-linear user--item interactions beyond matrix factorization, but performance typically depends on interaction density. Graph-based recommenders such as LightGCN \cite{ref13} and self-supervised methods \cite{ref14} address sparsity via structural priors or contrastive objectives, yet healthcare recommendation demands conservative claims due to the gap between offline metrics and clinical outcomes \cite{ref8}. To inject auxiliary supervision under extreme sparsity without requiring dense historical interactions, we introduce \textbf{Pseudo-Label Neural Collaborative Filtering (PL-NCF)}: we augment MF, MLP, and NeuMF with a soft pseudo-label objective derived from survey--group feature alignment. Crucially, PL-NCF maintains dual embedding spaces: main embeddings for ranking and PL-specific embeddings for semantic clustering, enabling specialization for distinct tasks.

This dual-representation approach relates to multi-task learning \cite{ref30,ref31}, where shared representations benefit from joint optimization of related tasks. However, rather than sharing all parameters, PL-NCF maintains separate embedding spaces for each objective, allowing task-specific specialization. Beyond top-$K$ ranking metrics, we evaluate embedding geometry using spherical $k$-means, cosine silhouette scores \cite{ref29}, and 2D visualizations using t-SNE \cite{ref28_tsne}. To avoid an unfair baseline in embedding analysis, we apply the same $k$-selection protocol to all models and representations: for each model, seed, entity, and representation, we evaluate $k \in \{3, 4, 5, 6, 7, 8, 10\}$ in the original embedding space and select the $k$ that maximizes cosine silhouette with tie-break to the smallest $k$. Crucially, we observe a separability--accuracy trade-off: higher main-embedding clusterability negatively correlates with ranking performance (Spearman $\rho \approx -0.38$ on leave-one-out; $\rho \approx -0.59$ on the train-validation-test split), suggesting that embeddings optimized for ranking may sacrifice interpretability.

\paragraph{Research questions.}
\begin{itemize}
  \item \textbf{RQ1:} Do survey-derived soft pseudo-labels improve NCF ranking performance (HR@5, NDCG@5) under extreme interaction sparsity, and how do effects vary across MF, MLP, and NeuMF architectures?
  \item \textbf{RQ2:} Do PL models dedicated PL-specific embeddings exhibit superior clusterability compared to baseline main embeddings and PL models own main embeddings, supporting the dual-representation hypothesis?
  \item \textbf{RQ3:} Is the optimal number of embedding clusters $k$ model- and representation-dependent, and does fixed $k = 5$ mischaracterize embedding structure?
  \item \textbf{RQ4:} What is the relationship between embedding clusterability and ranking accuracy in sparse-data regimes?
\end{itemize}

\paragraph{Contributions.}
\begin{enumerate}
  \item \textbf{Dual-representation PL-NCF framework.} We extend MF, MLP, and NeuMF with an auxiliary feature-alignment objective that learns separate PL-specific embeddings alongside main ranking-optimized embeddings, enabling task-specialized representations.
  \item \textbf{Fair cluster-count selection for embedding analysis.} We show that the optimal number of clusters is model- and representation-dependent, and we apply the same $k$-selection procedure to all models to avoid biased comparisons.
  \item \textbf{Empirical analysis on sparse OHC data.} We show that PL variants improve leave-one-out ranking metrics and that PL-specific embeddings achieve higher cosine silhouette scores than baseline main embeddings under optimal-$k$ spherical clustering.
  \item \textbf{Separability--accuracy trade-off.} We observe a negative correlation between main-embedding clusterability at fixed $k = 5$ and ranking performance, providing empirical evidence that interpretability and recommendation accuracy can compete under sparsity.
\end{enumerate}

\section{Related Work}
\subsection{OHC support and group formation}
OHCs have been shown to support peer-to-peer sense-making and reduce isolation \cite{ref1,ref2}, with evidence linking participation to improved self-management \cite{ref3}. Prior work on healthcare group formation has leveraged the textual content of discussions, such as topic modeling \cite{ref27}. Our focus differs: we assume onboarding-time access to structured survey features and study how to recommend support groups when interaction logs are sparse or minimal.

\subsection{Group recommender systems and fairness}
Group recommender systems typically aggregate individual preferences to recommend items to a group \cite{ref4,ref5,ref6}. Fairness-aware group recommendation has been studied to avoid marginalizing less active members \cite{ref9,ref10}. In our setting, we recommend groups to individuals in sparse-data contexts; fairness remains important for healthcare deployment, but our work primarily addresses sparse supervision and representation learning.

\subsection{Neural collaborative filtering and sparse-data recommendation}
NCF \cite{ref11} introduced GMF, MLP, and NeuMF to model non-linear interactions, building on matrix-factorization foundations \cite{ref32,ref33}. Graph-based recommenders such as NGCF \cite{ref12} and LightGCN \cite{ref13} further exploit connectivity structure, but performance often depends on interaction density. Sparse-data and cold-start recommendation can incorporate side information and self-supervised or contrastive objectives \cite{ref7,ref14}. In healthcare, conservative framing is critical because offline metrics do not necessarily translate to clinical outcomes \cite{ref8}. We leverage structured surveys to provide a feature-alignment pseudo-label signal that augments sparse interaction supervision.

\subsection{Pseudo-labeling and multi-task learning}
Pseudo-labeling \cite{ref15} is a simple semi-supervised learning approach; it is related to entropy minimization \cite{ref16} and has been extended via curriculum strategies \cite{ref17} and analyses of confirmation bias \cite{ref18}. Our use of pseudo-labels differs from self-training: we derive fixed soft targets from survey--group alignment, avoiding feedback loops from a model labeling its own predictions.

Multi-task learning \cite{ref30,ref31} jointly optimizes related objectives to improve generalization. Multi-task recommender systems \cite{ref34} have shown benefits from combining ranking with auxiliary signals. Our dual-representation approach maintains separate embedding spaces for each task, enabling stronger task-specific specialization than fully shared representations.

\subsection{Healthcare recommender systems}
Healthcare recommender systems have been applied to rehabilitation and mental-health interventions \cite{ref19,ref20}. Recent work has explored language-model-based methods for mental-health support and conversational recommendation with an emphasis on safety \cite{ref21,ref22}. Our work contributes a survey-driven sparse-data framework for recommending support groups in OHCs, with conservative claims rooted in available offline signals.

\subsection{Embedding analysis and visualization}
Silhouette analysis \cite{ref29} provides a quantitative measure of cluster separation, while t-SNE \cite{ref28_tsne} and UMAP \cite{ref28} offer qualitative 2D visualizations of embedding geometry. We use 2D projections only for visualization, and we quantify cluster structure in the original embedding space to avoid distortion artifacts from non-linear dimensionality reduction.

\section{Methodology}

\subsection{Problem formulation}
Let $\mathcal{U}=\{u_1,\dots,u_n\}$ be a set of $n$ users and $\mathcal{G}=\{g_1,\dots,g_m\}$ be a set of $m$ support groups. Each user $u\in\mathcal{U}$ has a survey feature vector $\mathbf{x}_u\in\mathbb{R}^{16}$ derived from intake questionnaire responses, and each group $g\in\mathcal{G}$ has a feature profile $\mathbf{z}_g\in\mathbb{R}^{16}$ aggregated from member surveys (construction detailed in \Cref{sec:dataset}). Observed user--group memberships form an implicit-feedback matrix $\mathbf{Y}\in\{0,1\}^{n\times m}$, where $y_{ug}=1$ indicates observed membership.

Our goal is to learn a scoring function $s_\theta(u,g)\in[0,1]$ that ranks groups for each user under extreme sparsity (three observed memberships per user). The survey features $\mathbf{x}_u$ and $\mathbf{z}_g$ are always available, enabling feature-driven prediction even with minimal interaction history.

\subsection{Survey representations and AlignFeatures pseudo-labels}
\label{sec:alignfeatures}

\paragraph{User features.}
Each user 16-dimensional vector $\mathbf{x}_u$ concatenates two survey components: (i) six Q33 weights capturing support-preference dimensions and (ii) ten Q26 weights reflecting demographic and health-condition factors. Weights within each component are normalized to sum to 1, yielding a unit-simplex representation per component.

\paragraph{Group features.}
Each group profile $\mathbf{z}_g\in\mathbb{R}^{16}$ is constructed via $k$-nearest-neighbor aggregation with $k=6$ of member survey features, with three repetitions per user during dataset generation. This yields 498 groups for 165 users (details in \Cref{sec:dataset}).

\paragraph{AlignFeatures pseudo-label.}
For each observed user--group pair $(u,g)$ with $y_{ug}=1$, the dataset provides a feature-alignment score:
\begin{equation}
\text{Align}(u,g) = \frac{\cos(\mathbf{x}_u, \mathbf{z}_g) + 1}{2} \in [0,1],
\end{equation}
mapping cosine similarity to the unit interval. We treat $\tilde{y}_{ug}=\text{Align}(u,g)$ as a fixed soft target during training, motivated by a homophily assumption common in peer-support settings: similarity in structured needs and context, as captured by intake surveys, is a reasonable proxy for short-list suitability when behavioral evidence is absent.

\subsection{PL-NCF architectures}
We compare three NCF architectures \cite{ref11}---Matrix Factorization (MF), Multi-Layer Perceptron (MLP), and NeuMF---each with a baseline variant (binary supervision only) and a pseudo-label (PL) variant.

\paragraph{Baseline models.}
\begin{itemize}
  \item \textbf{MF}: $\hat{y}_{ug} = \sigma(\mathbf{p}_u^\top \mathbf{q}_g)$, where $\mathbf{p}_u,\mathbf{q}_g\in\mathbb{R}^{d_{\text{MF}}}$ and $\sigma$ is the sigmoid.
  \item \textbf{MLP}: $\hat{y}_{ug} = \text{MLP}([\mathbf{p}_u; \mathbf{q}_g])$, with $\mathbf{p}_u,\mathbf{q}_g\in\mathbb{R}^{d_{\text{MLP}}}$.
  \item \textbf{NeuMF}: $\hat{y}_{ug} = \sigma(\mathbf{w}^\top[\mathbf{p}_u^{\text{GMF}}\odot\mathbf{q}_g^{\text{GMF}}; \text{MLP}([\mathbf{p}_u^{\text{MLP}};\mathbf{q}_g^{\text{MLP}}])])$.
\end{itemize}

\paragraph{PL models with dual embedding spaces.}
PL variants introduce a second set of embeddings:
\begin{itemize}
  \item \textbf{Main embeddings} $\mathbf{p}_u, \mathbf{q}_g$ optimized for interaction prediction.
  \item \textbf{PL-specific embeddings} $\mathbf{p}_u^{\text{PL}}, \mathbf{q}_g^{\text{PL}} \in \mathbb{R}^{d_{\text{PL}}}$ dedicated to learning feature-alignment structure.
\end{itemize}
The PL branch computes:
\begin{equation}
a_{ug}^{\text{PL}} = \cos(\mathbf{p}_u^{\text{PL}}, \mathbf{q}_g^{\text{PL}}) = \frac{\mathbf{p}_u^{\text{PL}\top}\mathbf{q}_g^{\text{PL}}}{\|\mathbf{p}_u^{\text{PL}}\|_2\|\mathbf{q}_g^{\text{PL}}\|_2},
\end{equation}
which is fused with the main prediction pathway (architecture-dependent fusion). For MF-PL and NeuMF-PL, we additionally project $\mathbf{x}_u$ to dimension $d_{\text{PL}}$ and compute cosine similarity against $\mathbf{q}_g^{\text{PL}}$ as an auxiliary signal.

\begin{figure*}[t]
  \centering
  \includegraphics[width=0.85\textwidth]{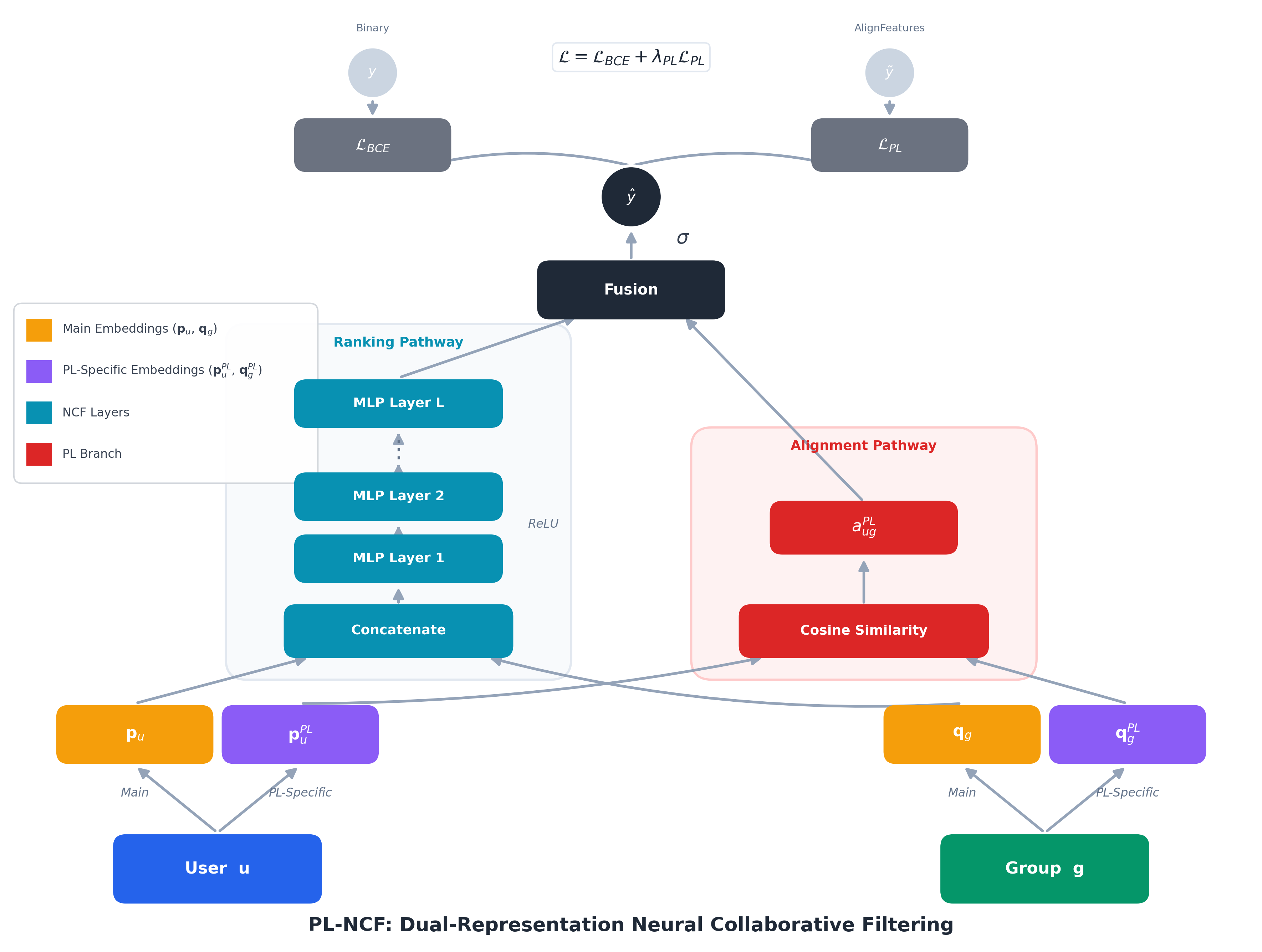}
  \caption{Overview of the PL-NCF dual-representation architecture. Each user $u$ and group $g$ maintain separate main embeddings ($\mathbf{p}_u$, $\mathbf{q}_g$) for the ranking pathway and PL-specific embeddings ($\mathbf{p}_u^{\text{PL}}$, $\mathbf{q}_g^{\text{PL}}$) for the alignment pathway. The ranking pathway processes main embeddings through NCF layers, while the alignment pathway computes cosine similarity $a_{ug}^{\text{PL}}$ between PL-specific embeddings. Both pathways are fused and supervised jointly by binary cross-entropy $\mathcal{L}_{\text{BCE}}$ and pseudo-label loss $\mathcal{L}_{\text{PL}}$.}
  \Description{Architecture diagram of PL-NCF showing dual embedding spaces, ranking pathway with MLP layers, alignment pathway with cosine similarity, and the combined training objective.}
  \label{fig:architecture}
\end{figure*}

\paragraph{Architectural details.}
We use standard NCF-style architectures with model-specific embedding sizes.
MF baselines use $d_{\text{MF}}=64$; MF-PL uses $d_{\text{MF}}=96$ for main MF embeddings and $d_{\text{PL}}=32$ for the PL branch.
MLP uses $d_{\text{MLP}}=32$ and a single hidden layer of size 32.
NeuMF uses $d_{\text{GMF}}=32$, $d_{\text{MLP}}=64$, and MLP layers $\{128, 64, 32\}$; PL variants use $d_{\text{PL}}=32$. \Cref{fig:architecture} illustrates the full PL-NCF dual-representation architecture.

\subsection{Training objective}
We train with on-the-fly uniform negative sampling: for each observed positive $(u,g)$ with $y_{ug}=1$, we sample one negative $g'\sim\mathcal{G}\setminus\{g' \mid y_{ug'}=1\}$. The baseline loss is binary cross-entropy:
\begin{equation}
\mathcal{L}_{\text{BCE}} = -\sum_{(u,g)\in\mathcal{D}} \big[y_{ug}\log\hat{y}_{ug} + (1-y_{ug})\log(1-\hat{y}_{ug})\big],
\end{equation}
where $\mathcal{D}$ is the training set including sampled negatives.

For PL models, when AlignFeatures $\tilde{y}_{ug}$ is available for pair $(u,g)$, we add a soft-label cross-entropy term:
\begin{equation}
\mathcal{L}_{\text{PL}} = -\sum_{(u,g)\in\mathcal{D}_{\text{PL}}} \big[\tilde{y}_{ug}\log\hat{y}_{ug} + (1-\tilde{y}_{ug})\log(1-\hat{y}_{ug})\big],
\end{equation}
where $\mathcal{D}_{\text{PL}}\subseteq\mathcal{D}$ contains pairs with AlignFeatures scores. The combined objective is:
\begin{equation}
\mathcal{L} = \mathcal{L}_{\text{BCE}} + \lambda_{\text{PL}}\mathcal{L}_{\text{PL}},
\end{equation}
where $\lambda_{\text{PL}}$ controls pseudo-label supervision strength.

\subsection{Embedding clustering and visualization}
After training, we extract user and group embedding matrices. For PL models, we extract both main embeddings and PL-specific embeddings.

\paragraph{Spherical $k$-means clustering in high-dimensional space.}
We apply $k$-means clustering to $\ell_2$-normalized embeddings (spherical $k$-means) \cite{ref29}. For each embedding set, we evaluate $k \in \{3, 4, 5, 6, 7, 8, 10\}$ and compute cosine silhouette scores using cosine distance on normalized embeddings. The silhouette computation is performed in the original embedding space, not on any 2D projection. For each model, seed, entity, and representation, we select the optimal $k$ (the $k$ maximizing cosine silhouette; tie-break: smallest $k$), then report mean and standard deviation across seeds. For the separability--accuracy analysis in \Cref{sec:sepacc}, we additionally report main-embedding silhouette at fixed $k=5$ to avoid post-hoc tuning effects.

\paragraph{2D visualization using t-SNE.}
We project high-dimensional embeddings to 2D using t-SNE \cite{ref28_tsne} for qualitative inspection. We apply $\ell_2$ normalization before projection and use cosine distance with perplexity 15. Crucially, spherical $k$-means cluster labels and silhouette metrics are computed in the original high-dimensional embedding space; the resulting high-dimensional cluster labels are then overlaid on the 2D t-SNE coordinates.

\section{Experimental Setup}

\subsection{Dataset}
\label{sec:dataset}
We use a survey-driven support-group dataset constructed from OHC intake questionnaires. The dataset contains:
\begin{itemize}
  \item \textbf{Users}: $n=165$, each with a 16-dimensional survey vector $\mathbf{x}_u$.
  \item \textbf{Support groups}: $m=498$, each with a 16-dimensional aggregated feature profile $\mathbf{z}_g$ generated via $k$-nearest-neighbor aggregation with $k=6$ of member surveys with three repetitions per user.
  \item \textbf{Observed memberships}: 498 total (three per user), representing synthetic group assignments based on feature similarity.
\end{itemize}
This synthetic construction reflects a bootstrapping scenario where groups are created from user needs rather than organic community formation.

\subsection{Evaluation protocols}
We evaluate under two data-split regimes:
\begin{enumerate}
  \item \textbf{Leave-one-out (primary)}: For each user, one interaction is held out for validation and one for test; remaining interactions (approximately one per user) form the training set. This protocol best reflects the extreme cold-start regime central to our work.
  \item \textbf{Train-validation-test split (70/15/15)}: A random split over all 498 observed memberships, providing supplementary analysis under slightly less extreme sparsity.
\end{enumerate}

\subsection{Metrics}

\paragraph{Ranking metrics.}
We evaluate top-$K$ ranking with $K=5$ using sampled evaluation: for each held-out positive, we sample 99 negatives uniformly from groups not observed in training for that user, forming a candidate set of 100. We report HR@5 and NDCG@5. Metrics are aggregated over test users and reported as percentages.

\paragraph{Clustering metrics.}
We quantify embedding clusterability via cosine silhouette score computed on $\ell_2$-normalized embeddings with cosine distance. For each model, seed, entity, and representation, we select $k \in \{3, 4, 5, 6, 7, 8, 10\}$ maximizing cosine silhouette, then report mean and standard deviation across seeds (\Cref{tab:silhouette}). For the separability--accuracy analysis, we use main-embedding silhouette at fixed $k=5$.

\subsection{Implementation details}
Models are implemented in PyTorch and trained for 20 epochs with AdamW. We report mean and standard deviation over five random seeds (42, 52, 62, 122, 232). Hyperparameters follow the architecture-specific settings in \Cref{sec:alignfeatures} through \Cref{sec:dataset}, with model- and protocol-specific $\lambda_{\text{PL}}$ values (MF-PL: 0.03 leave-one-out and 0.40 train-validation-test split; MLP-PL: 0.25 and 0.20; NeuMF-PL: 0.35 and 0.50). For 2D visualizations, we use t-SNE with $\ell_2$-normalized embeddings, cosine distance, perplexity 15, and seed-specific random state. All clustering decisions and silhouette metrics are computed in the original embedding space and only visualized in 2D via overlay.

\section{Results}

\subsection{Ranking performance}
\Cref{tab:loo} reports our primary leave-one-out results, and \Cref{tab:stratified} provides supplementary results on the train-validation-test split.

\begin{table}[t]
\caption{Recommendation performance under leave-one-out evaluation (primary), mean \(\pm\) std over 5 seeds.}
\label{tab:loo}
\centering
\begin{tabular}{l l l l}
\toprule
Model & {HR@5 (\%)} & {NDCG@5 (\%)}\\
\midrule
MF       & 4.58 $\pm$ 1.93 & 2.70 $\pm$ 1.33\\
MF-PL    & \textbf{5.42 $\pm$ 2.25} & \textbf{3.32 $\pm$ 1.51} \\
MLP      & 2.65 $\pm$ 1.63 & 1.41 $\pm$ 0.97 \\
MLP-PL   & \textbf{5.30 $\pm$ 1.88} & \textbf{2.97 $\pm$ 1.12}\\
NeuMF    & 4.46 $\pm$ 0.91 & 2.50 $\pm$ 0.58 \\
NeuMF-PL & \textbf{5.18 $\pm$ 1.25} & \textbf{3.02 $\pm$ 0.63} \\
\bottomrule
\end{tabular}
\end{table}

\begin{table}[t]
\caption{Supplementary result: Recommendation performance on train-validation-test (70/15/15) split, mean \(\pm\) std over 5 seeds.}
\label{tab:stratified}
\centering
\begin{tabular}{l l l l}
\toprule
Model & {HR@5 (\%)} & {NDCG@5 (\%)} \\
\midrule
MF       & 4.57 $\pm$ 1.86 & 2.90 $\pm$ 1.47 \\
MF-PL    & 2.57 $\pm$ 1.86 & 2.15 $\pm$ 1.48\\
MLP      & 0.57 $\pm$ 0.78 & 0.23 $\pm$ 0.32\\
MLP-PL   & \textbf{1.43 $\pm$ 1.75} & \textbf{0.72 $\pm$ 1.04}\\
NeuMF    & 3.43 $\pm$ 0.78 & 2.43 $\pm$ 0.78\\
NeuMF-PL & \textbf{6.29 $\pm$ 0.78} & \textbf{3.90 $\pm$ 0.35}\\
\bottomrule
\end{tabular}
\end{table}

\subsection{Embedding clustering quality}
\Cref{tab:silhouette} reports cosine silhouette scores for user and group embeddings under leave-one-out. PL-specific embeddings achieve higher clusterability than both baseline embeddings and PL models own main embeddings.

\begin{table*}[t]
\caption{Cosine silhouette scores under leave-one-out evaluation (mean \(\pm\) std over 5 seeds). Each entry uses per-seed optimal \(k \in \{3, 4, 5, 6, 7, 8, 10\}\) computed in the original embedding space.}
\label{tab:silhouette}
\centering
\begin{tabular}{l l l l l}
\toprule
Model & {User Main} & {User PL} & {Group Main} & {Group PL} \\
\midrule
MF       & 0.0394 $\pm$ 0.0018 & -- & 0.0318 $\pm$ 0.0007 & -- \\
MF-PL    & 0.0265 $\pm$ 0.0020 & \textbf{0.0684 $\pm$ 0.0050} & 0.0223 $\pm$ 0.0013 & \textbf{0.0572 $\pm$ 0.0011} \\
MLP      & 0.0687 $\pm$ 0.0013 & -- & 0.0577 $\pm$ 0.0015 & -- \\
MLP-PL   & 0.0680 $\pm$ 0.0026 & \textbf{0.0716 $\pm$ 0.0028} & 0.0569 $\pm$ 0.0017 & \textbf{0.0567 $\pm$ 0.0015} \\
NeuMF    & 0.0263 $\pm$ 0.0018 & -- & 0.0222 $\pm$ 0.0008 & -- \\
NeuMF-PL & 0.0256 $\pm$ 0.0011 & \textbf{0.0653 $\pm$ 0.0022} & 0.0220 $\pm$ 0.0004 & \textbf{0.0571 $\pm$ 0.0015} \\
\bottomrule
\end{tabular}
\end{table*}

\subsection{Separability--accuracy trade-off}
\label{sec:sepacc}
We analyze the relationship between main-embedding silhouette at fixed $k=5$ and ranking performance across model--seed runs. Spearman rank correlation \cite{ref24} reveals a negative relationship:
\begin{itemize}
  \item \textbf{Leave-one-out (primary)}: $\rho_{\text{sil, HR@5}} \approx -0.38$
  \item \textbf{Train-validation-test split (supplementary)}: $\rho_{\text{sil, HR@5}} \approx -0.59$
\end{itemize}
This indicates that models with more clusterable main embeddings tend to achieve lower ranking accuracy in this sparse-data setting.

\subsection{2D visualization using t-SNE}
\begin{figure*}[t]
  \centering
  \includegraphics[width=0.48\textwidth]{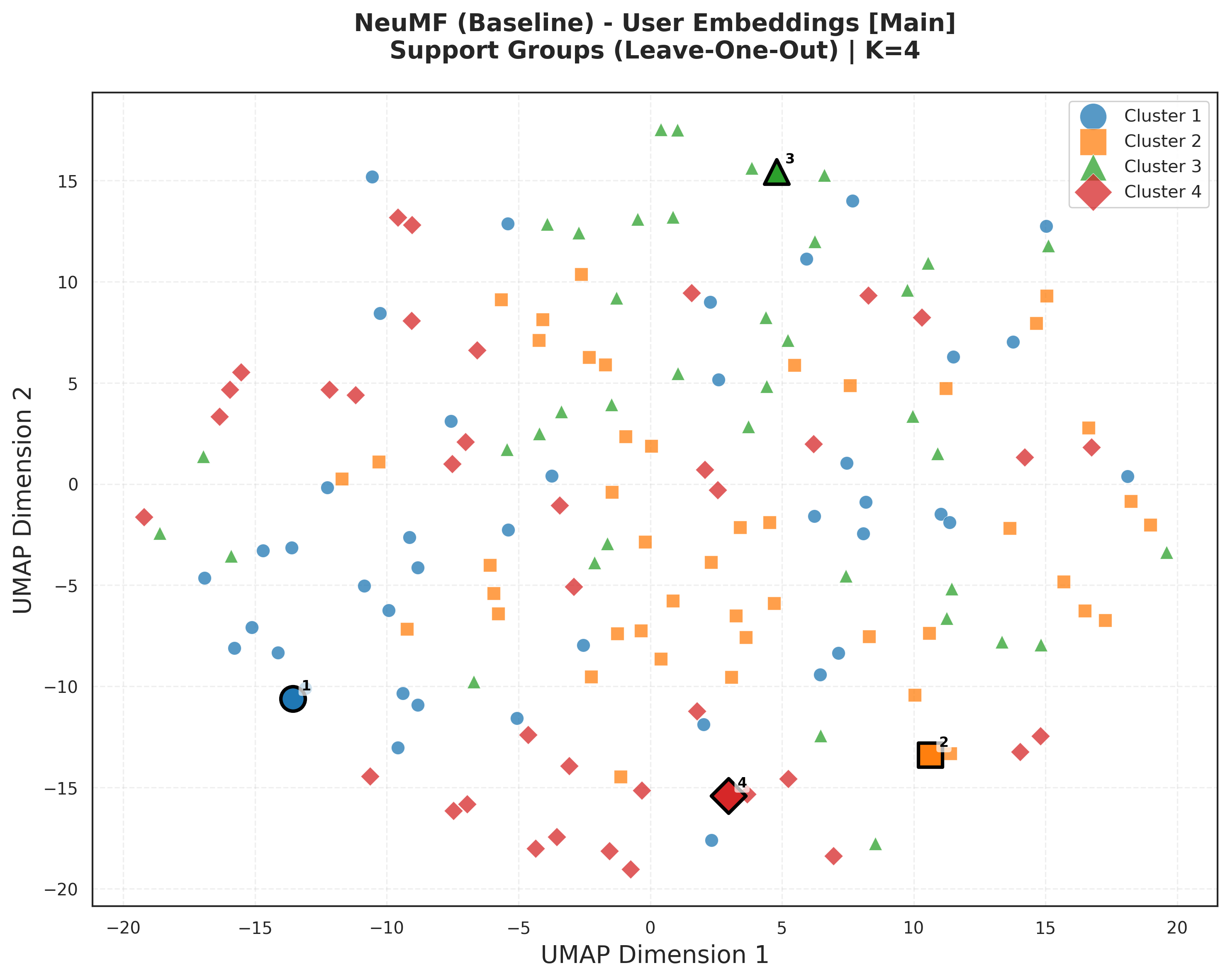}
  \includegraphics[width=0.48\textwidth]{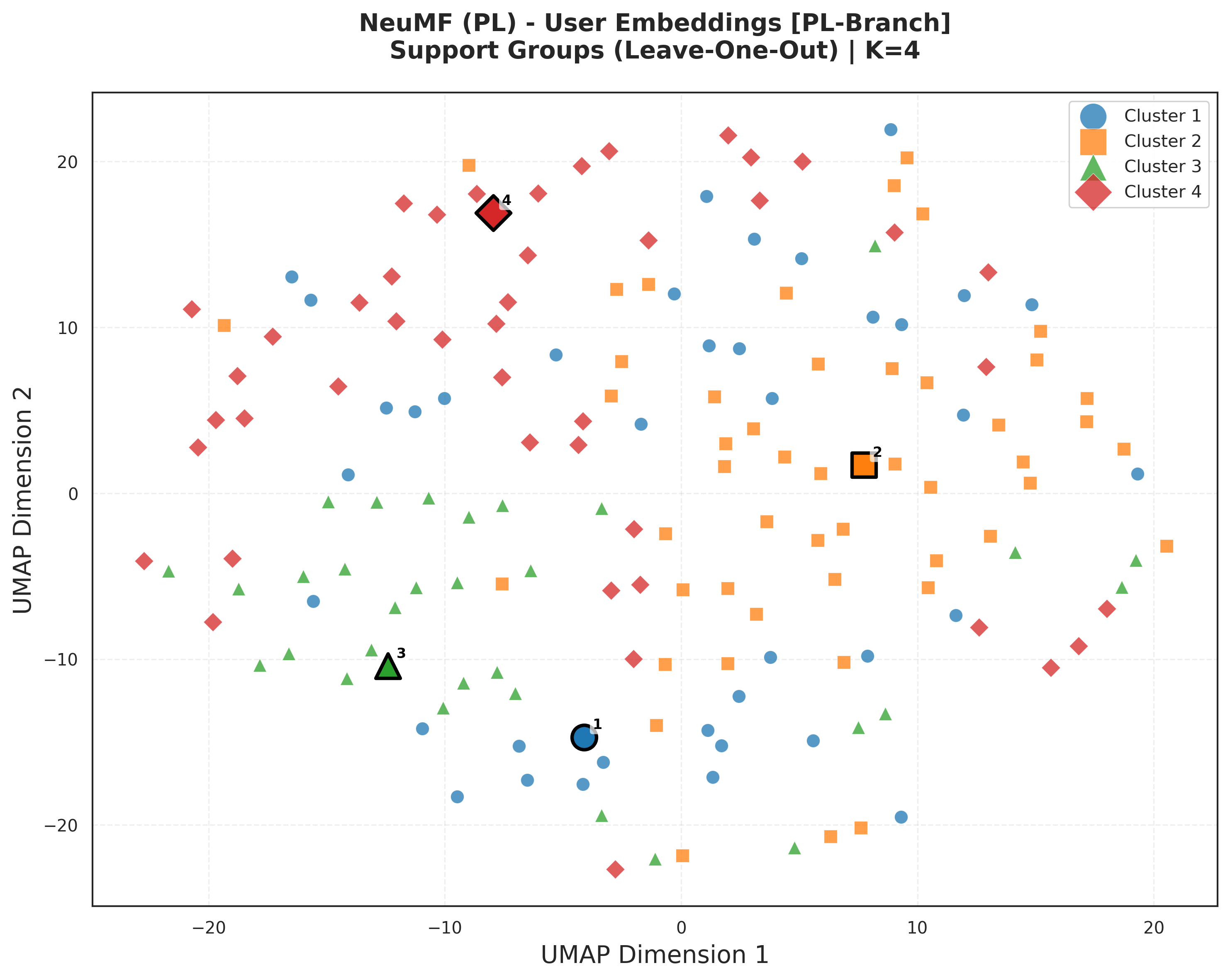}
  \caption{t-SNE visualization of user embeddings under leave-one-out evaluation. Left: NeuMF baseline main embeddings. Right: NeuMF-PL PL-specific embeddings. Cluster labels are computed via spherical $k$-means in the original embedding space and overlaid on the 2D coordinates for visualization only.}
  \Description{Two side-by-side t-SNE scatter plots of user embeddings under leave-one-out evaluation, colored by high-dimensional spherical k-means cluster assignment.}
  \label{fig:tsne_neumf_baseline_vs_pl}
\end{figure*}

\Cref{fig:tsne_neumf_baseline_vs_pl} shows an example t-SNE projection of user embeddings under leave-one-out comparing NeuMF baseline main embeddings to NeuMF-PL PL-specific embeddings. Points are colored by spherical $k$-means cluster labels computed in the original embedding space, with $k$ selected by the same high-dimensional silhouette procedure.

\section{Discussion}

\subsection{When does pseudo-labeling improve ranking}
Pseudo-labeling exhibits protocol- and architecture-dependent effects. Under leave-one-out---our primary evaluation reflecting extreme sparsity with approximately one training positive per user---all three PL variants consistently improve both HR@5 and NDCG@5, demonstrating that pseudo-label regularization is most beneficial when collaborative signals are minimal. This consistency across all architectures is a key finding. Under the train-validation-test split, gains are architecture-dependent: NeuMF-PL achieves large improvements, while MF-PL degrades. This divergence suggests that when more training data is available, hybrid architectures can leverage auxiliary signals constructively, whereas bilinear models may experience interference from pseudo-label objectives that conflict with interaction-structure learning.

\subsection{Why do PL-specific embeddings cluster better}
Three mechanisms explain PL-specific embedding improvements:
\begin{enumerate}
  \item \textbf{Objective decoupling}: main embeddings optimize ranking loss; PL-specific embeddings are shaped by feature-alignment supervision. Separate embedding spaces enable task-specific specialization.
  \item \textbf{Cosine-consistent geometry}: AlignFeatures derives from cosine similarity, and we evaluate clustering using spherical $k$-means with cosine silhouette, aligning supervision and evaluation geometry.
  \item \textbf{Feature-grounded semantics}: PL-specific embeddings receive direct supervision from survey-derived similarities, encoding interpretable feature-based groupings independent of interaction structure.
\end{enumerate}

\subsection{Implications for healthcare recommendation}
Survey-driven recommendation is pragmatic in healthcare: intake questionnaires are standard practice, and structured features avoid privacy-sensitive text mining. However, AlignFeatures is not ground-truth preference; offline metric improvements do not guarantee engagement or clinical outcomes without real-world validation. Our dataset is small and synthetically constructed via neighbor aggregation, limiting generalizability. We position this work as a proof-of-concept for dual-representation learning under extreme sparsity, not a deployable clinical system.

\subsection{Broader implications for representation learning}
The negative correlation between main-embedding clusterability and ranking accuracy suggests caution when interpreting intrinsic embedding metrics or visually appealing 2D projections as proxies for downstream performance. Dual-representation architectures provide a practical mechanism to obtain task-specialized embeddings when interpretability and ranking compete.

\section{Limitations}
\paragraph{Feature-similarity proxy for preference.}
AlignFeatures pseudo-labels are derived from survey--group feature similarity, not observed user engagement or satisfaction. Improvements in offline HR@5 under sampled evaluation do not directly imply improved real-world engagement or health outcomes.

\paragraph{Small-scale synthetic dataset.}
The dataset contains 165 users and 498 support groups with three memberships per user, constructed via nearest-neighbor aggregation of survey features rather than organic community formation. The small scale limits statistical power and increases variance across seeds.

\paragraph{2D projections are qualitative only.}
t-SNE provides intuitive 2D visualizations of embedding geometry but involves non-linear dimensionality reduction with sensitivity to hyperparameters and random seeds. We therefore compute clustering and silhouette scores in the original embedding space and use 2D projections only for qualitative visualization via label overlays.

\paragraph{Limited baseline breadth.}
We compare PL-augmented versus baseline NCF variants but do not include non-NCF baselines such as purely feature-based ranking or recent self-supervised approaches. These omissions limit claims about broader state-of-the-art performance.

\section{Conclusion}
We presented PL-NCF, a survey-driven pseudo-labeling framework for Neural Collaborative Filtering in sparse-data support-group recommendation. This dual-representation approach maintains separate main embeddings optimized for ranking and PL-specific embeddings shaped by feature-alignment pseudo-labels, enabling task specialization. Under leave-one-out evaluation---our primary protocol reflecting extreme cold-start sparsity---pseudo-labeling consistently improves ranking across all three architectures, while train-validation-test split results are architecture-dependent. In the leave-one-out setting, PL-specific embedding spaces exhibit higher clusterability than baseline main embeddings under optimal-$k$ spherical clustering, and we observe a separability--accuracy trade-off: embedding clusterability can be negatively correlated with ranking quality. Future work must validate survey-derived alignment signals against real OHC engagement and outcomes and explore richer survey encoders under conservative and fairness-aware evaluation.

\section{Reproducibility}
We release code, configuration, and derived artifacts needed to reproduce the reported results, including trained model checkpoints across five seeds, extracted embedding matrices, per-$k$ silhouette grids, and 2D visualization figures. We emphasize that all clustering decisions and cluster quality metrics are computed in the original embedding space; 2D projections are used only to visualize those high-dimensional cluster labels.

\bibliographystyle{ACM-Reference-Format}
\bibliography{bibliography}

\appendix
\section{Additional Embedding Visualizations}
\label{sec:appendix_visualizations}
\begin{figure*}[t]
  \centering
  \begin{tabular}{cc}
    \includegraphics[width=0.48\linewidth]{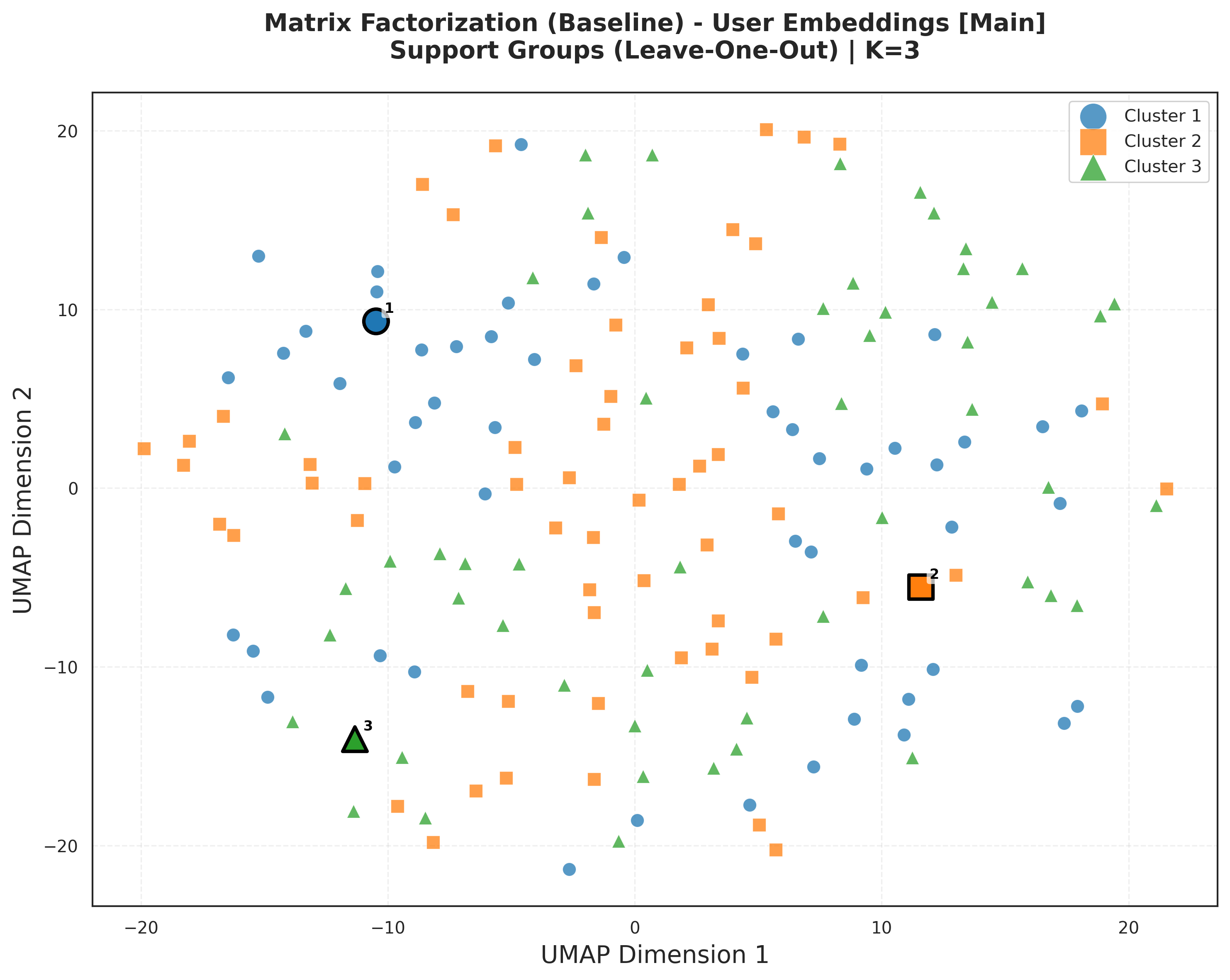} &
    \includegraphics[width=0.48\linewidth]{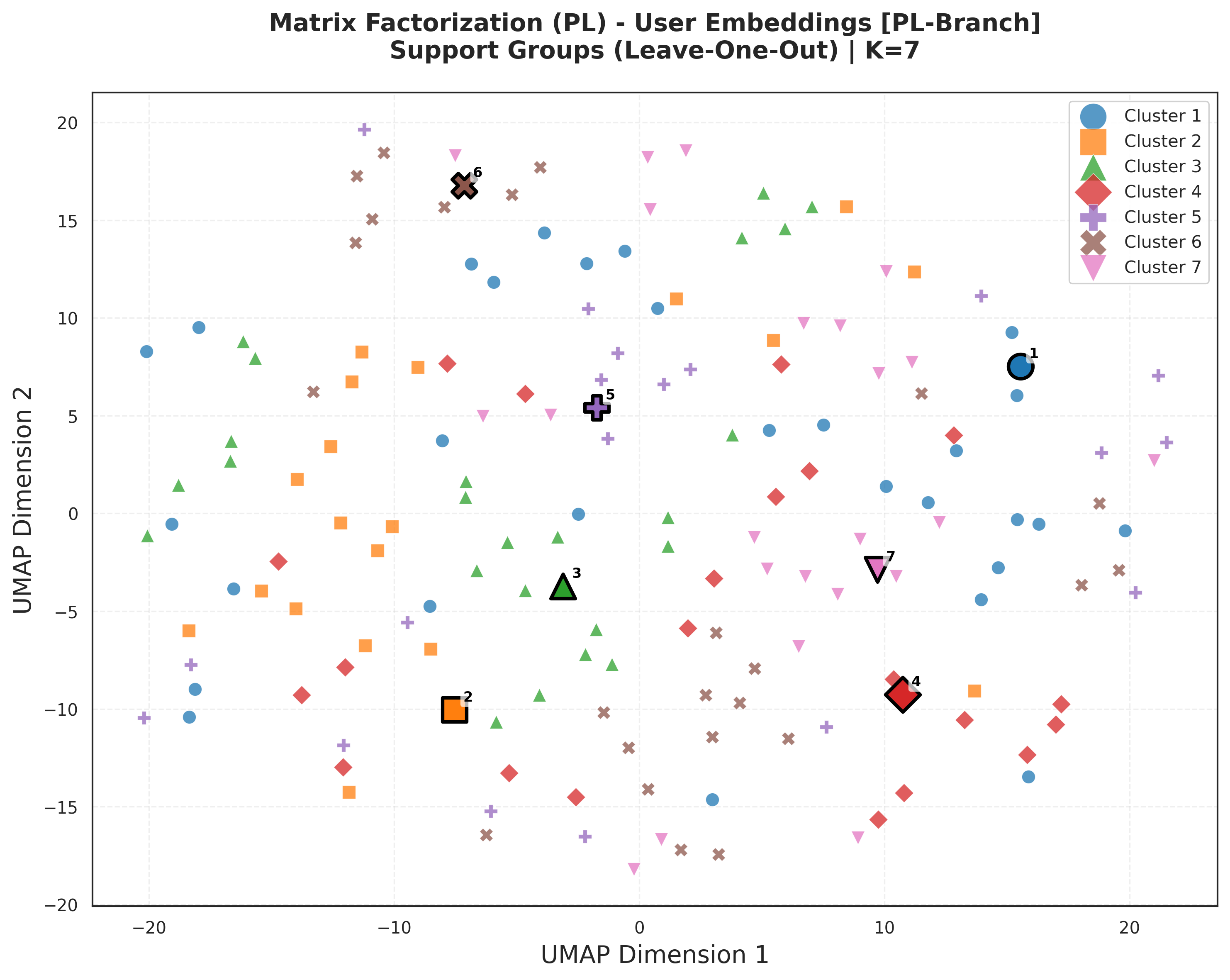} \\
    \small (a) MF baseline main embeddings & \small (b) MF-PL PL-specific embeddings \\[1em]
    \includegraphics[width=0.48\linewidth]{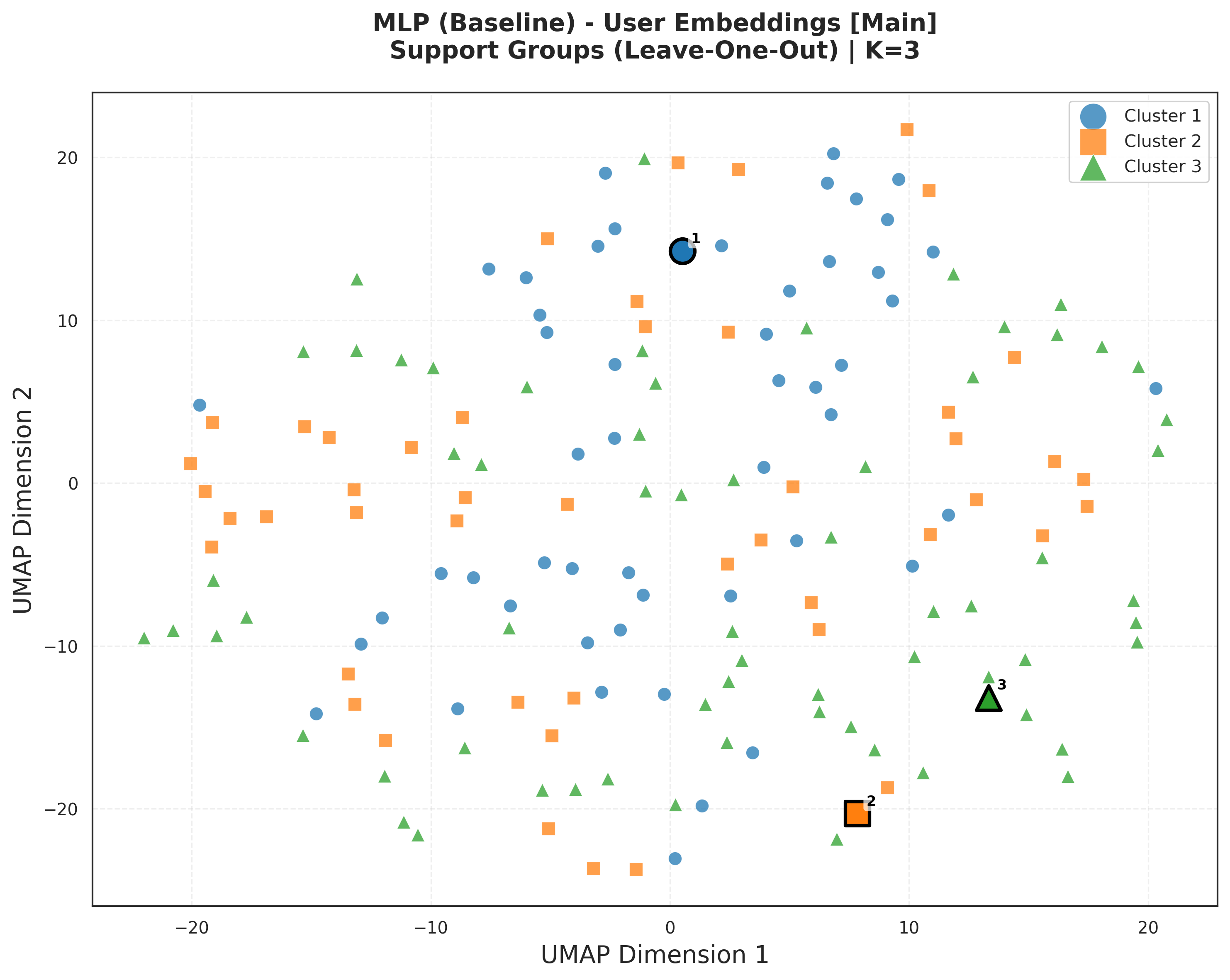} &
    \includegraphics[width=0.48\linewidth]{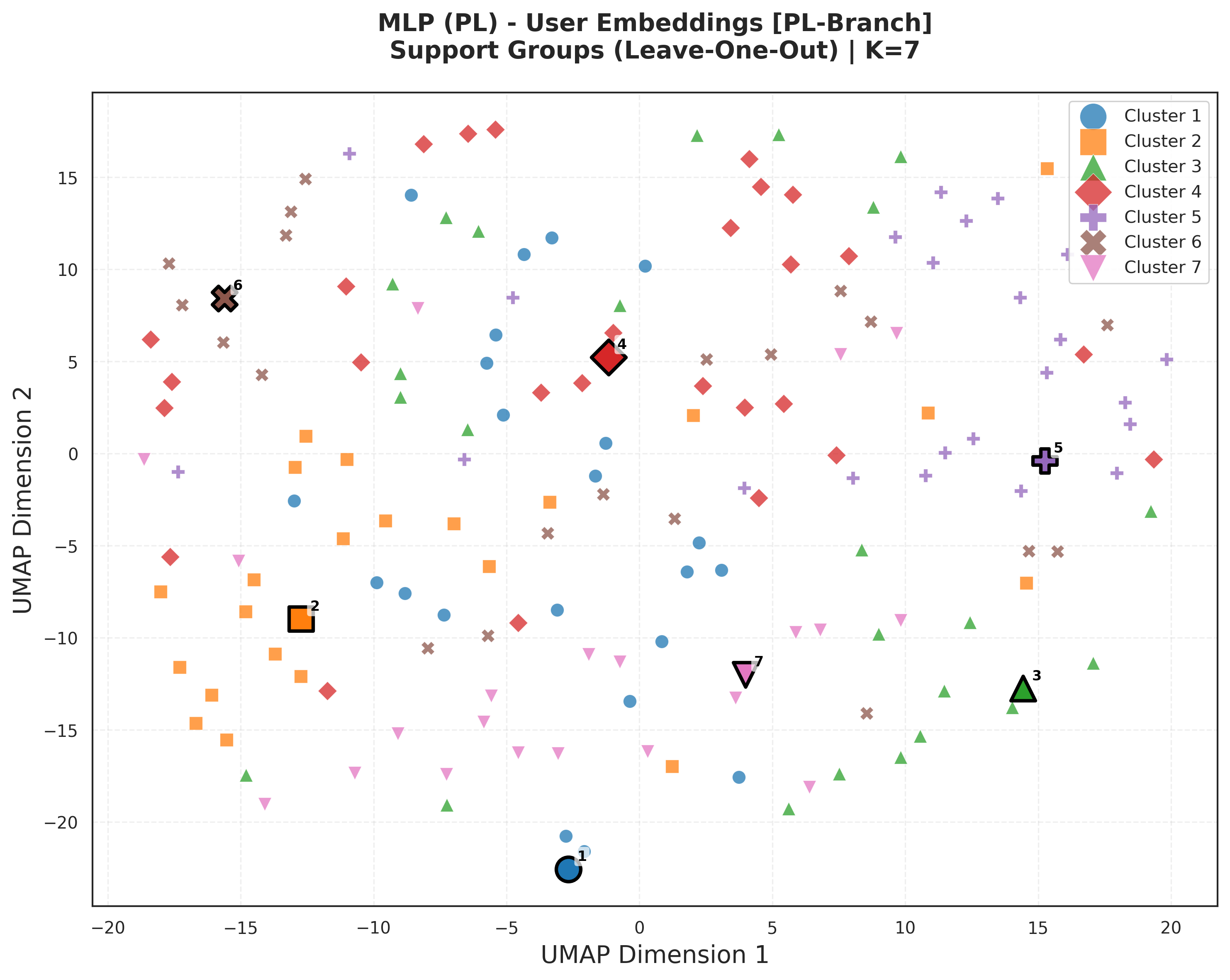} \\
    \small (c) MLP baseline main embeddings & \small (d) MLP-PL PL-specific embeddings
  \end{tabular}
  \caption{Additional t-SNE visualizations under leave-one-out comparing baseline main embeddings (left) to PL-specific embeddings (right) for MF and MLP. Cluster labels are computed via spherical $k$-means in the original embedding space and overlaid on 2D projections for visualization only.}
  \Description{Four t-SNE scatter plots arranged in 2 rows and 2 columns, comparing baseline main embeddings and PL-specific embeddings for MF and MLP.}
  \label{fig:tsne_appendix_mf_mlp}
\end{figure*}

\Cref{fig:tsne_appendix_mf_mlp} provides additional t-SNE visualizations under leave-one-out comparing baseline main embeddings to PL-specific embeddings for MF and MLP. These figures are included for qualitative inspection and are not used as inputs to clustering or evaluation.

\end{document}